\documentclass[oupdraft]{bio}
\usepackage{dsfont}
\usepackage{url}
\usepackage{color}
\definecolor{mediumgray}{gray}{0.625}
\usepackage{adjustbox}
\usepackage{multirow, makecell}
\usepackage[normalem]{ulem}
\usepackage{placeins}
\usepackage{comment}
\usepackage[]{multibib}

\newcites{supplement}{References for Supplement}

\newcommand{\adAbbrev}{\textsc{ad}}
\newcommand{\biocard}{\textsc{biocard}}
\newcommand{\apoe}{\textsc{apoe}{\small 4}}
\newcommand{\Apoe}{\textsc{Apoe}{\small 4}}
\newcommand{\csf}{\textsc{csf}}
\newcommand{\mri}{\textsc{mri}}

\newcommand{\smooth}{\mathrm{s}}
\newcommand{\vanishderiv}{\mathrm{v}}
\newcommand{\observed}{\mathrm{obs}}
\newcommand{\randomeffect}{\mathrm{rnd}}
\newcommand{\indFun}{\mathds{1}}
\newcommand{\normalDist}{\mathcal{N}}

\begin{document}
	
\title{Bayesian shape-constrained regression for quantifying Alzheimer's disease biomarker progression}

\author{MINGYUAN LI$^1$, ZHEYU WANG$^{1, 2, \ast}$, AKIHIKO NISHIMURA$^1$\\[4pt]
	\textit{$^1$Department of Biostatistics, Johns Hopkins University, Baltimore, Maryland, United States\\
    $^2$ Division of Quantitative Sciences, Sidney Kimmel Comprehensive Cancer Center, Johns Hopkins University, Baltimore, Maryland, United States}
	}

\markboth
{M. Li and others}
{Bayesian shape-constrained regression for Alzheimer's}

\maketitle

\footnotetext{To whom correspondence should be addressed: wangzy@jhu.edu}

\begin{abstract}{
		Several biomarkers are hypothesized to indicate early stages of Alzheimer’s disease, well before the cognitive symptoms manifest. 
		Their precise relations to the disease progression, however, is poorly understood. 
		This lack of understanding limits our ability to diagnose the disease and intervene effectively at early stages.
		To provide better understanding of the relation between the disease and biomarker progressions, we propose a novel modeling approach to quantify the biomarkers' trajectories as functions of age.
		Building on monotone regression splines, we introduce two additional shape constraints to incorporate structures informed by the current medical literature.
		First, we impose the regression curves to satisfy a vanishing derivative condition, reflecting the observation that changes in biomarkers generally plateau at early and late stages of the disease. 
		Second, we enforce the regression curves to have a unique inflection point, which enhances interpretability of the estimated disease progression and facilitates assessment of temporal ordering among the biomarkers.
		We fit our shape-constrained regression model under Bayesian framework to take advantage of its ability to account for the heterogeneity in disease progression among individuals.
		When applied to the \biocard{} data, the model is able to capture asymmetry in the biomarkers' progressions while maintaining interpretability, yielding estimates of the curves with temporal ordering consistent with the existing scientific hypotheses.
}
{Bayesian inference, hierarchical model; regression splines; longitudinal analysis; neurodegenerative disorders}

\end{abstract}
	
\section{Introduction}
\label{s:intro}

Multiple biomarkers are used in clinical practice and research related to Alzheimer's disease (\adAbbrev{}) as indicators of the disease state. These biomarkers include deposition of Amyloid substance, 
neurodegeneration, and decline in cognitive performance, reflecting different aspects of \adAbbrev{} pathophysiological processes.
Better quantification of the relationship between these biomarkers and disease progression can provide opportunities for early diagnosis and effective intervention before irreversible brain structural changes occur (\citealp{Gauthier2015, Wang2020}).

The Jack model is a widely adopted scientific hypothesis characterizing the \adAbbrev{} biomarker progression (\citealp{jack2010, Jack2013}). 
The model summarizes existing bodies of evidence into two key features.
The first feature is the assumption of certain temporal ordering in the progression of \adAbbrev{} biomarkers. 
For example, the abnormality in cerebrospinal fluid biomarkers is believed to precede brain structure changes, which in turn precede cognitive deterioration. 
The model's second feature is the non-linear evolution of \adAbbrev{} biomarkers, where changes in biomarkers are smaller at early and late clinical stages while greater in-between.
It is additionally believed that substantial heterogeneity in the biomarker progression exists among individuals.

Multiple parametric statistical models have been proposed for describing the trajectories of \adAbbrev{} biomarkers. 
\citet{Jedynak2012} assume the population-level trajectory of each biomarker to be a logistic function $f(t;c, s, h)=\frac{h}{1+\exp\left( -\frac{t-c}{s} \right)}$, with $t$ representing a latent disease progression level and $(c, s, h)$ unknown parameters to be estimated.
\citet{Li2017} transform biomarker measurements to empirical quantiles and then, through inverse Gaussian cumulative distribution function transform, to a real line.
They then model these transformed biomarkers as a linear function of age. 
With these parametric methods, there are concerns of model misspecifications and of the resulting biased estimates producing an incorrect temporal ordering of biomarkers.

Semi- and non-parametric regression models provide more flexible alternatives that avoid potential biases from misspecification. 
In the context of \adAbbrev{} biomarker modeling, \citet{Li2018} use a second-degree polynomial spline regression to model the association of cognitive biomarkers to time since disease onset among dementia patients. 
\citet{AntonianoVillalobos2014} use a Bayesian non-parametric mixture model to study change over time in hippocampal volume, a common marker of \adAbbrev{}. 
Their model uses an infinite mixture of parametric regression models, similar to kernel density estimation, to characterize the distribution of biomarkers as a function of age and other covariates.

The unconstrained nature of these semi- and non-parametric models has its drawbacks, however, with their estimates susceptible to outliers and other idiosyncrasies of observational data.
For instance, \citet{AntonianoVillalobos2014} find that, according to their estimate, the hippocampal volume at population level decreases over time but rebounds starting around 80 years of age. 
Such sudden recovery is physiologically implausible and is likely an artifact of noises. 
More generally, the short follow-up and low measurement frequency make it difficult to fit overly flexible models to \adAbbrev{} datasets. 
Many longitudinal studies on \adAbbrev{} have at most 5 biomarker measures per individual and take fewer than 1 measurement per year (\citealp{Lawrence2017}).

In this article, we propose a shape-constrained spline regression model that strikes a balance between robustness and flexibility.
Our model incorporate scientifically-informed structures into the regression curves while retaining semi-parametric flexibility. 
We place a set of linear inequality constraints on the spline coefficients to ensure the estimated curve to be monotone and have  a unique inflection point.
We additionally impose gradual flattening of the regression curve towards early and late stages of life by placing increasingly strong penalty on magnitudes of spline coefficients towards the boundary knots.
To fit this model under Bayesian framework, we express the shape-constraints in the form of parameter constraints and informative priors. 
The Bayesian framework allows us to straightforwardly incorporate hierarchical structures into the model and account for the high-degree of heterogeneity among individuals in the \adAbbrev{} application. 

The rest of the article is structured as follows.
Section~\ref{bayes-ssr} introduces our shape-constraint regression model and describes how to fit it under the Bayesian paradigm. 
Section~\ref{s:method} describes additional components necessary in modeling the longitudinal \adAbbrev{} biomarker data. 
Section~\ref{s:simulation} compares the performances of our shape-constrained model against the parametric and unconstrained model through a simulation study. 
Section~\ref{s:result} applies our model to the \biocard{} data and discusses the results of this analysis.
Section~\ref{s:discuss} concludes with discussions for future research directions for \adAbbrev{} biomarker modeling.

\section{Shape-constrained Spline Regression}\label{bayes-ssr}

\begin{figure}[htp]
	\includegraphics[width=.9\linewidth]{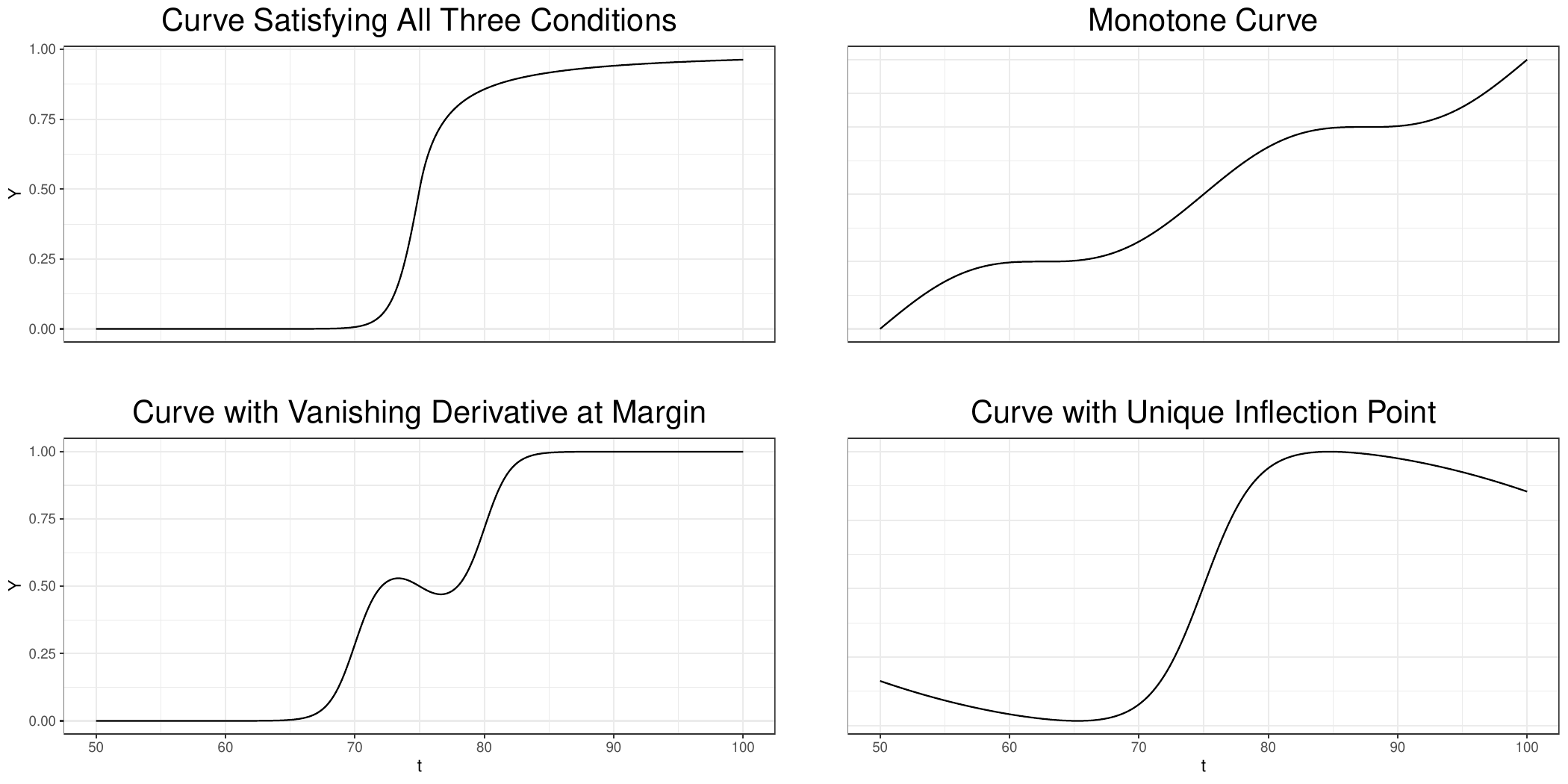}
	\caption{
		Example of curves satisfying either the monotonicity, vanishing derivative, or unique inflection property. 
		The top left curve satisfy all the three and others only one of the properties.}
	\label{curve-example}
\end{figure}

In this section, we present shape-constrained regression as a general technique of independent interest, deferring to Section~\ref{s:method} modeling aspects specific to the \adAbbrev{} application. 
We describe methods to impose three properties on the estimated curve---monotonicity, unique inflection point, and vanishing derivative---through appropriate constraints on the spline coefficients. 
Figure~\ref{curve-example} shows examples of three curves, each satisfying only one of the three properties.
Also shown in the figure is a curve satisfying all the three properties; 
we will refer to such curves as \textit{S-shaped} for the rest of the article. 
After introducing the shape-constraints, we discuss how to fit the model under Bayesian paradigm.
The Bayesian approach is useful in \adAbbrev{} applications, where the substantial heterogeneity among individuals warrants a large number of patient-level random effects. 

\subsection{Review of spline regression}
\label{spline-review}

\newcommand{\splineIndex}{m}
\newcommand{\numSplines}{M}
\newcommand{\splineBasis}{B}
Spline regression provides a flexible way to model a predictor $t$'s relation to the outcome $Y$. 
The idea is to model the expected value of $Y$ as a non-linear function of $t\in[L,U]$ as
\begin{equation*}
f(t;\gamma) = \sum_{\splineIndex = 1}^\numSplines \gamma_\splineIndex \splineBasis_\splineIndex(t),
\end{equation*}
where $\splineBasis_1(t), \ldots, \splineBasis_\numSplines(t)$ are pre-specified basis functions and $\gamma=(\gamma_1,\ldots,\gamma_\numSplines)$ are unknown coefficients to be estimated \citep{wegman1983splines}.
A common choice of the basis is quadratic B-splines, which are piecewise quadratic polynomials defined on an interval $[L,U]$ with a series of knot points $L=\zeta_1<\zeta_2< \dots <\zeta_{\numSplines-2} < \zeta_{\numSplines-1}=U$.
For $3\le \splineIndex\le \numSplines-2$, the $\splineIndex$-th basis function is supported on $[\zeta_{\splineIndex-2}, \zeta_{\splineIndex+1}]$ and is continuously differentiable at both ends of the support. 
The boundary basis functions behave slightly differently;
the first and second basis functions are supported on $[L,\zeta_2]$ and $[L,\zeta_3]$ and have non-zero derivatives at $t = L$.
Similarly, the $(\numSplines-1)$-th and $\numSplines$-th basis functions are supported on $[\zeta_{\numSplines-3},U]$ and $[\zeta_{\numSplines-2},U]$ and have non-zero derivatives at $t = U$.
These four boundary basis functions control the boundary behavior of $f(t;\gamma)$.

Spline models become more flexible as the number of basis functions increases, but so does their tendency to overfit. 
Penalized spline regression is a common approach to avoid overfitting while still retaining flexibility \citep{eilers1996psplines}. 
The method adds roughness penalty to the negative log-likelihood, encoding preference towards smoother curve estimates.
More precisely, the method estimates the regression curve by minimizing the penalized negative log-likelihood:
\begin{equation}\label{general-penalty-LL}
-l_\mathrm{pen}(\gamma; y, \sigma_\smooth)=-l(\gamma; y)+S_\smooth(\gamma; \sigma_\smooth),
\end{equation}
where $-l(\gamma; y)$ is the negative log-likelihood, given for example by
$$-l(\gamma; y) = \sigma_\observed^{-2} \sum_{i, j} \left(y_{i}(t_{ij}) - f(t_{ij}; \gamma) \right)^2$$
when assuming a model $\, y_i(t) = f(t; \gamma) + \epsilon_i$ for $\epsilon_i \sim \mathcal{N}\! \left(0, \sigma_\observed^2 \right)$, 
and $S_\smooth(\gamma; \sigma_\smooth)$ is roughness penalty on $f(t; \gamma)$ with penalty strength controlled by the hyperparameter $\sigma_\smooth$.  
One common choice of $S_\smooth(\gamma; \sigma_\smooth)$ penalizes larger differences between neighboring spline coefficients $\gamma_\splineIndex$ and $\gamma_{\splineIndex + 1}$ as
\begin{equation}\label{smooth-penalty}
	S_\smooth(\gamma; \sigma_\smooth) = \sigma_\smooth^{-2} \sum_{\splineIndex=1}^{\numSplines-1} (\gamma_{\splineIndex+1} - \gamma_{\splineIndex})^2.
\end{equation}

\subsection{Monotonicity}
\label{mono-assumption}

In applications such as modeling of \adAbbrev{} biomarker progression, it is natural from scientific perspectives to assume the function $f(t; \gamma)$ to be monotonic.
This is equivalent to assuming its derivative to have a constant sign so that either $f'(t; \gamma) \geq 0$ or $f'(t; \gamma) \leq 0$ for all $t \in [L, U]$. 
Under the spline regression framework, one way to achieve this is to model $f'(t; \gamma)$ in terms of B-splines and require the coefficients $\gamma_\splineIndex$'s to be all non-negative or all non-positive. 
This gives rise to monotone spline regression (\citealp{ramsay1988monotone_regression}):
\begin{equation*} 
	f(t)=\sum\limits_{\splineIndex=1}^\numSplines \gamma_{\splineIndex} I_\splineIndex(t),
\end{equation*}
where $I_\splineIndex(t) = \int_L^t B_\splineIndex(s) \, \textrm{d} s$ is the \textit{integrated spline} basis and the coefficients are constrained to be either $\gamma_\splineIndex \geq 0$ or $\gamma_\splineIndex \leq 0$ for all $m = 1, \ldots, \numSplines$.
In our \adAbbrev{} application, we assume the biomarker values to represent abnormality so that it is natural to assume $\gamma_m \geq 0$ and $f(t; \gamma)$ to be increasing.
Other than the monotonicity constraint, the model retains the flexibility of spline regression through its modeling of $f'(t; \gamma)$ with B-spline basis.

\subsection{Unique inflection point}
\label{s-assumption}
Here we present another type of a shape constraint on spline regression, which can be imposed independently or along with the monotonicity constraint. 
An inflection point of $f(t; \gamma)$ is where its derivative achieves a local maximum or minimum.
When we expect the existence of a ``turning point'' in the predictor $t$'s relation to the outcome, toward which the rate of change increases and beyond which it decreases, we might require the curve to have a unique inflection point. 
As we are particularly interested in estimating $f(t; \gamma)$ as a monotone increasing function, we focus on the case where the inflection point corresponds to a local maximum of $f'(t;\gamma)$.

To incorporate the unique-inflection constraint into spline regression, we observe that it is equivalent to unimodality of the derivative $f'(t;\gamma)$.
Following \citet{gunn2005unimodal} and \citet{Kllmann2014}, we impose the unimodality through linear inequalities
\begin{equation}\label{umb}
	\gamma_{1}\le\gamma_{2}\le\cdots\le
	\gamma_{\splineIndex^* - 1} \leq \gamma_{\splineIndex^*} \ge \gamma_{\splineIndex^*+1}
	\ge\cdots\ge\gamma_{\numSplines}
	\ \text{ for some } \,
	\splineIndex^* \in \{1,2,\cdots,\numSplines\}.
\end{equation}
We refer to the index $\splineIndex^*$ as the \textit{inflection index} of $\gamma$.
Since $f'(t;\gamma)$ is a linear combination of B-splines, the inequalities \eqref{umb} on $\gamma$ implies $f'(t;\gamma)$ to be monotone on $[L,U]$ and achieve a unique maximum at some $t^*\in[L, U]$ \citep{Kllmann2014}, which is also the inflection point of $f(t;\gamma)$ . 
When using the piecewise cubic integrated splines, the inflection index $\splineIndex^*$ is related to the actual inflection point $t^*$ through the fact that $\,t^*\in[\zeta_{\splineIndex^*-2},\zeta_{\splineIndex^*-1}]$ if $\,2\le\splineIndex^*\le \numSplines-1$.
And $f'(t;\gamma)$ will be monotonically decreasing or increasing on $[L, U]$  when $\splineIndex^*=1\text{ or }\numSplines$. 

\subsection{Vanishing derivative}
\label{vanish-deriv-assumption}
We now introduce yet another type of constraint, \textit{vanishing derivative} condition, for spline regression.
This constraint is motivated by the fact that, in our \adAbbrev{} application, the biomarkers are expected to remain stable and low before disease onsets and reach a plateau toward the end of the disease.
More generally, the constraint is of interest whenever we expect an analogous relation between an outcome and predictor and wish to assume $f'(t;\gamma) \to 0$ as $t \to L$ or $t \to U$. 

In the context of spline regression, the vanishing derivative property amounts to requiring the coefficients of the boundary basis functions to vanish; i.e.\ $\gamma_1 = \gamma_2 = \gamma_{\numSplines-1} =\gamma_\numSplines = 0$.
Additionally, it is reasonable to impose a gradual change in the derivative through a soft constraint of the form $0 \approx |\gamma_3| \lesssim |\gamma_4| \lesssim |\gamma_5| \lesssim \ldots \gtrsim |\gamma_{\numSplines-3}| \gtrsim |\gamma_{\numSplines-2}| \approx 0$ on the spline coefficients near the boundaries. 
This soft constraint is achieved by increasingly strong penalty on the magnitude of $\gamma_m$'s towards the boundaries.  
We can express such penalty on $\gamma$ as
\begin{equation} \label{vanish-deriv-penalty}
	S_\vanishderiv(\gamma; \sigma_\vanishderiv^2) 
	\propto \frac{1}{\sigma_\vanishderiv^2} \sum\limits_{\splineIndex=1}^{\numSplines}\frac{\gamma_{\splineIndex}^2}{\alpha_\numSplines(\splineIndex)},
\end{equation}
where $\alpha_\numSplines(\splineIndex) \geq 0$ is a symmetric \textit{window function} that starts zero at both ends, with $\alpha_\numSplines(\splineIndex) = 0$ for $\splineIndex \in \{1, 2, \numSplines - 1, \numSplines\}$, and gradually rises to $\alpha_\numSplines(\splineIndex) = 1$ towards the middle;
with slight abuse of notation, we use $\alpha_\numSplines(\splineIndex) = 0$ to imply infinity penalty $1 / \alpha_\numSplines(m) = \infty$ which enforces a hard constraint $\gamma_\splineIndex = 0$. 
The hyperparameter $\sigma_\vanishderiv^{-2}$ additionally controls the strength of penalty for $\gamma_\splineIndex$ with $\alpha_\numSplines(\splineIndex) > 0$. 

While other choices are possible, here we use a Planck-taper window function~(\citealp{McKechan_2010}).
Setting $m' = m + 2$ and $M' = M - 4$ to focus on the non-boundary spline coefficients, this window function is defined as
\begin{equation*}
	\alpha_\numSplines(\splineIndex')=\begin{cases}
		0 & \text{ for }\, \splineIndex' \leq 0 \\
		\left(1+\exp\left(\frac{0.1\numSplines'}{\splineIndex'}-\frac{0.1\numSplines'}{\numSplines' - \splineIndex'}\right)\right)^{-1} 
			& \text{ for }\, 1\le \splineIndex' < 0.1 \numSplines'\\
		1 & \text{ for }\, 0.1\numSplines' \le \splineIndex' \le \numSplines' / 2
	\end{cases}
\end{equation*}
and $\alpha_\numSplines(\splineIndex')$ for $\splineIndex' > \numSplines' / 2$ are defined in a symmetric manner.

\subsection{Bayesian approach to shape-constrained regression}
\label{bayes-smooth}

We now describe how to fit our shape-constrained spline model under Bayesian framework. 
Earlier works have considered Bayesian modeling of regression curves with monotonicity or unimodal constraint in isolation \citep{Neelon2004, gunn2005unimodal, Katja2009}.
The application of unimodality constraint to the derivative, to achieve the unique-inflection property, is a distinguishing feature of our work. 
The combination of the unique-inflection property with monotonicity and the introduction of vanishing derivatives further distinguish our work. 

We start by observing that penalizing a negative log-likelihood has a Bayesian interpretation as placing a prior distribution on unknown parameters.
In particular, minimizing the penalized negative log-likelihood of the form \eqref{general-penalty-LL} is equivalent to maximizing 
$$
\exp\!\big\{
	l_\textrm{pen}(\gamma; y, \sigma_\smooth)) 
\big\}
	= L(\gamma; y) \exp(- S_\smooth(\gamma; \sigma_\smooth) ),
$$
where the right hand side can be interpreted as, up to a normalizing constant, the posterior distribution under the prior $\propto \exp(- S_\smooth(\gamma; \sigma_\smooth) )$ on $\gamma$ (\citealp{miller2019bayesian}). 
More generally, we can combine the penalties \eqref{smooth-penalty} for smoothness and \eqref{vanish-deriv-penalty} for vanishing derivative into an unnormalized prior 
\begin{equation*}
	\exp (
		- S_\smooth(\gamma; \sigma_\smooth) - S_\vanishderiv(\gamma; \sigma_\vanishderiv) 
	)
	= \exp\!\left(
		- \frac{1}{\sigma_\smooth^2} \sum_{\splineIndex=2}^\numSplines (\gamma_{\splineIndex} - \gamma_{\splineIndex-1})^2
		- \frac{1}{\sigma_\vanishderiv^2} \sum_{\splineIndex=1}^{\numSplines} \frac{\gamma_{\splineIndex}^2}{\alpha_\numSplines(\splineIndex)}
	\right),
\end{equation*}
From this Bayesian perspective, the hyperparameter $\sigma_\smooth$ can be thought of as an average scale of change in successive spline coefficients, the hyperparameter $\sigma_\vanishderiv$ an average size of spline coefficients, and the window function $\alpha_\numSplines(\splineIndex)$ relative prior variances on $\gamma_{\splineIndex}$'s.

The additional monotonicity constraint $\gamma_\splineIndex \geq 0$ and unique-inflection constraint \eqref{umb} together amount to truncating the prior distribution to a region $\Gamma = \cup_{\splineIndex^* = 1}^\numSplines \Gamma_{\splineIndex^*} \subset \mathbb{R}^\numSplines$ where
\begin{equation}
\label{eq:prior_constraint}
\Gamma_{\splineIndex^*} = \big\{
(\gamma_1,\gamma_2,\cdots,\gamma_{\numSplines}): 
\, 0 \leq \gamma_{1} \le \ldots \le
	\gamma_{\splineIndex^* - 1} \leq \gamma_{\splineIndex^*} \ge \gamma_{\splineIndex^*+1}
	\ge\ldots\ge\gamma_{\numSplines} \geq 0
\big\}.
\end{equation}
This truncation combined with the smoothness and vanishing derivative penalties translates into the following prior on $\gamma$ for our Bayesian shape-constrained regression model 
\begin{equation*} 
p(\gamma \, | \, \sigma_\smooth^2, \sigma_\vanishderiv^2)
	\propto \exp\!\left(
		- \frac{1}{\sigma_\smooth^2} \sum_{\splineIndex=2}^\numSplines (\gamma_{\splineIndex} - \gamma_{\splineIndex-1})^2
		- \frac{1}{\sigma_\vanishderiv^2} \sum_{\splineIndex=1}^{\numSplines} \frac{\gamma_{\splineIndex}^2}{\alpha_\numSplines(\splineIndex)}
	\right) \indFun\{ \gamma \in \Gamma \}.
\end{equation*}
The hyperparameters $\sigma_\smooth$ and $\sigma_\vanishderiv$ can also be estimated under Bayesian framework by placing a hyperprior on them, such as a weakly informative half-normal prior.

The Bayesian framework allows straightforward quantification of uncertainty in the estimates of regression curves as well as of their functionals by mapping the posterior of $\gamma$ into that of appropriate quantities of interest. 
This is one major advantage of our Bayesian approach; 
while $\gamma$ may be estimated through constrained optimization under frequentist paradigm, it would be more challenging to derive confidence intervals.

\section{Multivariate Longitudinal Model for AD Biomarkers}
\label{s:method}

We now turn to building a model for \adAbbrev{} biomarker data, a critical component of which is the shape-constrained spline method developed in the previous section. 
We focus on the model structure here, deferring details on the \adAbbrev{} biomarkers and baseline covariates to Section~\ref{data-desc}.

\subsection{Bayesian hierarchical modeling of \adAbbrev{} biomarkers}\label{ss:hetero}

We denote the $i$-th subject's baseline characteristics by $x_{i} \in \mathbb{R}^q$ for $i = 1, \ldots, N$, age at the $j$-th by $t_{ij}$ for $j = 1, \ldots, J_i$, and the $k$-th biomarker value by $y_{ijk}$ for $k = 1, \ldots, K$.
We assume the signs of the biomarkers have been chosen so that higher biomarker values indicate more abnormality. 
We denote the population-level progressions of the biomarkers, the primary quantities of scientific interest, by $f_k(t; \gamma_k)$.

The baseline biomarker values vary significantly across individuals, some of which can be explained by their baseline characteristics but substantial portion of which cannot be.
To account for this heterogeneity, we hierarchically model the observed biomarker values with individual-level effects as follows: 
\begin{equation} \label{mean-model}
	\begin{split}
		y_{ijk} &= f_k(t_{ij}; \gamma_k)+x_{i}^T\beta_k+\omega_{ik}+\varepsilon_{ijk},\\
		\varepsilon_{ijk}&\overset{i.i.d.}{\sim}\normalDist(0,\sigma_\observed^2),\\
		\omega_{ik}&\overset{i.i.d.}{\sim}\normalDist(0,\sigma_\randomeffect^2),\\
	\end{split}
\end{equation}
where $\omega_{ik}$ represent the individual-level effects. 
We place a weak Gaussian prior with mean $\mu_\beta=0$ and variance $V_\beta=100^2 \mathrm{I}$ on the fixed effect $\beta$. 
On the variance parameters $\sigma_\observed^2$ and $\sigma_\randomeffect^2$, we place inverse-Gamma priors with shape $3$ and scale $0.5$ to assign a high prior probability to the interval $[0,1]$;
we make this informative choice since each biomarker outcome $y_k = \{ y_{ijk} \}_{ij}$ is standardized to have unit variance in our \adAbbrev{} application.
Also, variance parameters in a hierarchical model in general are difficult to identify and warrant informative priors \citep{gelman2006prior}.
The parametrization of and prior on $f_k(t; \gamma_k)$ are discussed in next section.

\subsection{Parametrization of biomarker progression curves}
\label{ss:model}

To incorporate the existing science of \adAbbrev{} mechanisms as summarized in the Jack model (Section~\ref{s:intro}), we model $f_k(t; \gamma_k)$ using the shape-constrained splines of Section~\ref{bayes-ssr}. 
Our interest in studying the biomarkers' temporal ordering makes the shape-constrained modeling particularly relevant. 
To characterize the relative ordering of the biomarkers, we rely on certain milestones in biomarker progressions.
One milestone of interest is the age at which each biomarker reaches the 50\% of maximum progression. 
Another milestone is the inflection point as it corresponds to the age at which the biomarker undergoes the fastest change.
The monotonicity and S-shape constraints ensure the existence of a unique 50\% threshold and of a unique inflection point, respectively.

Besides the shape constraints, we leverage the fact that the \adAbbrev{} biomarkers can be classified into three distinct categories based on their relations to underlying biological mechanisms (Section~\ref{s:result}).
Specifically, rather than modeling all the biomarkers independently, we assume the ones within the same category to share the same inflection index $\splineIndex^*$ (Equation~\ref{umb}), allowing them to borrow information from each other.

We parameterize $f_k(t; \gamma_k)$ using $\numSplines = 24$ basis functions to ensure sufficient flexibility.
Spline knot locations are commonly chosen as quantiles of the empirical distribution of observed data so that roughly the same amount of data is available for estimating each coefficient \citep{semiparaReg_2003}.
Such use of the empirical distribution, however, prevents us from inferring the biomarker progressions outside the observed age range of $\,t \in [50, 90]$.
To get around this issue, we smooth the empirical age distribution over range $[0, 120]$ through beta density kernels and choose the knot locations as $\numSplines$ quantiles of this smoothed distribution.
Details on the kernel density smoothing is provided in the supplement Section~\ref{beta-kde}.

On the hyperparameters $\sigma_\smooth^2$ and $\sigma_\vanishderiv^2$ for $\gamma_k$, we place a half-normal prior with scale $(\numSplines - 4)^{-1} = 1 / 20$, where $\numSplines - 4$ corresponds to the number of non-zero spline coefficients under the vanishing derivative constraint.
This scale is chosen because the biomarkers have been preprocessed to have unit variance;
we expect $f_k(120) - f_k(0)$, the change in the population-level progression over the lifespan, to have a similar scale.

\subsection{Posterior inference via Markov chain Monte Carlo}\label{MCMC}

To carry out inference under the model (\ref{mean-model}), we use the Gibbs sampling algorithm to draw from the posterior distribution of $(\beta, \gamma, \omega, \sigma_\observed^2,\sigma_\randomeffect^2,\sigma_\smooth^2,\sigma_\vanishderiv^2)$, where $\beta = \{ \beta_k \}_k$, $\gamma = \{\gamma_k\}_k$, and  $\omega = \{\omega_{ik}\}_{ik}$. 
The conditional update of $(\beta, \gamma)$ is done by first sampling the inflection indices $\splineIndex_k^*$ and then sampling $(\beta_k, \gamma_k)$ from truncated multivariate Gaussian distributions under the linearly constrained region $\Gamma_{\splineIndex_k^*}$ as in Equation \eqref{eq:prior_constraint}.
The sampling of $\splineIndex_k^*$ requires integration of a multivariate Gaussian density over $\Gamma_{\splineIndex^*}$ and this is done using the method of \cite{Genz1992}.
The subsequent sampling from truncated Gaussians are done using the algorithm of \cite{Pakman2014} as implemented in the \texttt{hdtg} package \citep{zhang2022hdtg}.
We update $(\sigma_\smooth^2, \sigma_\vanishderiv^2)$ using the Metropolis algorithm, the required density evaluations for which again require similar integration of a multivariate Gaussian density and use the method of \citet{Genz1992}.
The other conditional updates only involve standard parametric families. 
Further details on the Gibbs sampler is provided in the supplement Section \ref{mcmc-detail}.

We use the generated samples to calculate the posterior mean and $95\%$ credible intervals. 
For the progression curves $f_k(t) = f_k(t; \gamma_k)$, we construct point-wise credible intervals over $t \in [0, 120]$. 
When conducting inference on the biomarkers' temporal ordering, we account for the difference in the ranges of $f_k(t)$ by mapping them to the standardized scale as 
\begin{equation}
\label{eq:biomarker_standardization}
f_{k, \textrm{std}}(t)=\frac{f_k(t) - f_k(0)}{f_k(120) - f_k(0)} \in [0, 1].
\end{equation}
In both our simulation study of Section~\ref{s:simulation} and real-data application of Section~\ref{s:result}, we run the Gibbs sampler for 10,000 iterations, discard the first 5,000 as burn-in, and use the remaining 5,000 samples for inference.

\section{Simulation Study}
\label{s:simulation}

We conduct a simulation study to compare our S-shaped spline model against two alternatives. 
As the less flexible alternative, we consider a parametric model of the form $f(t;c, s, h)=\frac{h}{1+\exp\left( -\frac{t-c}{s} \right)}$ for the population-level progression curve. 
As the more flexible alternative, we consider a monotone spline model without the unique-inflection or vanishing derivative constraints.

We evaluate from multiple perspectives the models' abilities to estimate a true regression curve $f(t)$.
For each model, we calculate the mean square error (\textsc{mse}) of the posterior curve estimate $\hat{f}(t)$ at each time point $t$ and report the average of the pointwise \textsc{mse} over the time range. 
We similarly assess the frequentist coverage probability of the 95\% credible intervals by calculating the pointwise coverage and averaging it over the time range.
Besides the model's ability to estimate the pointwise values of $f(t)$, we also assess its ability to estimate the curve's inflection point of  as well as the 50\% threshold, i.e.\ the value of $t$ at which $f(t)$ reaches half of its maximum value. 
The inflection point and 50\% threshold both capture useful milestones of the biomarker progression and are of practical interests.
We evaluate these estimators again in terms of the \textsc{mse} and frequentist coverage.

For the S-shaped and unconstrained spline models, we also assess their sensitivities to the choice of the left and right boundaries of spline knots, which reflect the presumed earliest and latest possible times of disease progression.
Since we do not know when a biomarker actually starts or ceases to progress in the \adAbbrev{} application, it makes sense to choose the boundaries conservatively and ensure they cover the most likely range of actual disease progression.
On the other hand, choosing too wide a range likely leads to some loss in statistical efficiency.
To assess this potential sensitivity, we fit each model under two different choices of the spline boundaries, $\{30,90\}$ and $\{0,120\}$,
the former representing a range just wide enough to capture most of the change in $f$ and the latter a more conservative range.

\subsection{Simulation setup}
\label{sim:setup}

In order to assess performance of our method in a realistic scenario, we simulate synthetic datasets that emulate features of the \biocard{} data we analyze in Section~\ref{s:result}. 
In particular, our synthetic datasets mimic the \biocard{} data in sample size, presence of both binary and continuous baseline covariates, age distribution at baseline, number of observations for each individual, frequency of measurements, and high heterogeneity in baseline biomarker values.

For simplicity, we consider a model with a single biomarker. 
Given a true progression curve $f(t)$, which we specify momentarily, we simulate the observations according to the random-effect model \eqref{mean-model}. 
The sample size is chosen as $N = 250$.
The baseline covariates are simulated as $x_{i, 1}\sim\text{Bernoulli}(0.5)$ and $x_{i,2}\sim \normalDist(0,1)$.
We set the intercept as $\beta_0 = 0.4$ and the coefficients as $\beta_1 = -0.5$ and $\beta_2 = 0.1$. 
The age at first visit $t_{i1}$ is simulated from $\text{Uniform}(50,90)$ and the number of visits $n_i$ from $\text{Poisson}(10)$. 
The times between two consecutive visits $t_{i(j+1)}-t_{ij}$ are taken to be 1 and then slightly perturbed by adding $\operatorname{Exp}(\mathrm{scale} = 1/20)$. 
The baseline biomarker value $\omega_i$ is simulated from $\normalDist(0,1)$ and observation noise $\epsilon_{ij}$ from $\normalDist(0, 0.5)$. 
Finally, to emulate the fact that not all biomarkers are measured at each visit, we mark 30\% of the biomarker measurements as unobserved.

We consider two different choices for the true progression curve $f(t)$:
$$
\begin{aligned}
f_{\textrm{logit}}(t)&=\frac{2}{1+\exp(-\frac{t-70}{5})},\\
f_{\textrm{asym}}(t)&=\begin{cases}
	0 & \text{ for }t<30\\
	\frac{2(t-30)^3}{(75-30)^2(90-30)}  & \text{ for }30\le t<75\\
	2\left(1-\frac{(90-t)^3}{(90-75)^2(90-30)}\right)  & \text{ for }75\le t<90 \\
	2 & \text{ for }t\ge90.
\end{cases}
\end{aligned}
$$
The first function is symmetric around its inflection and 50\% threshold point $t = 70$. 
The second function, constructed as a piecewise cubic polynomial, is asymmetric and has distinct inflection and 50\% threshold points.
It reaches the 50\% of its overall height at $t=69.3$ but has an inflection point at $t = 75$, which also coincides with reaching the 75\% of its overall height.
The curves are chosen such that at least 98\% of the progression happen within interval $[30,90]$ and their derivatives essentially vanish at $0$ and $120$.
We simulate 100 datasets each under $f_{\textrm{logit}}$ and $f_{\textrm{asym}}$, with $x_i$, $t_{ij}$, and $y_{ij}$ independently generated for each dataset.

We fit the S-shaped spline model as specified in Section \ref{s:method}. 
For the more flexible spline model without the vanishing derivative and unimodality constraints, we modify the scale of the half-normal priors on $\sigma_\smooth^2,\sigma_\vanishderiv^2$ to  $0.01$ to keep the curve's prior expected height to be roughly 2. 
For the parametric model, we use the \texttt{rstan} package (\citealp{rstan_package}) to fit the model under weakly informative truncated Gaussian priors
$h\sim \normalDist^+(2, 1)$, 
$c\sim \normalDist^+(70,30^2)$ and 
$s\sim \normalDist^+(5, 1)$;
these priors are chosen to be centered around the truth for $f_{\textrm{logit}}$ to tilt the comparison in the parametric model's favor.

\subsection{Results}
\label{sim:results}

Table \ref{sim_result} summarizes the results of our simulation study. 
When the underlying truth is $f_{\textrm{logit}}$, the logistic model unsurprisingly has the smallest \textsc{mse}s and achieves coverage rates closest to the nominal 95\%. 
Without the parametric assumption, the S-shaped model is able to achieve performance much closer to the logistic model than to that of the more flexible spline model.
The S-shaped model is competitive with the logistic model in terms of \textsc{mse}s but somewhat lags behind in terms of coverage, with its credible intervals under-covering the progression curve itself and over-covering the progression milestones. 
The flexible model performs remarkably poorly; this behavior may be surprising at first but, as we explain below, has a clear root cause and reveals a major shortcoming of the unconstrained monotone spline model for the \adAbbrev{} application.

When the underlying truth is $f_{\textrm{asym}}$, the logistic model becomes mis-specified and its performance, while having comparable \textsc{mse} to the spline models, suffers from lower coverage.
The model's estimate of the inflection point exhibits a particularly severe bias with poor uncertainty quantification.
On the other hand, the S-shape model can better capture the asymmetry of the curve and emerges as a clear winner here.
The slight but noticeable degradation in its performance, compared to when the underlying truth is $f_{\textrm{logit}}$, is likely due to the fact that our prior is specified in a symmetric manner and that this leads to some degree of posterior bias in finite sample size.
The flexible model again performs poorly.

Under both truth settings, using the wider knot boundaries in the spline models tends to increase the \textsc{mse}s.
It also does not help with the over-coverage of the milestones observed when the true curve is logistic.
On the other hand, the wider knot boundaries helps the S-shaped model achieve coverage rates closer to the nominal 95\% when the true progression curve is asymmetric.
Since it is unlikely for a real progression curve to follow a specific symmetric parametric form, we recommend the more conservative, wider choice of the spline knot boundaries in practice to better account for uncertainties.

\begin{table}[tb]
	\centering
	\caption{
		Results of the simulation study comparing the parametric (``{Logistic}''), S-shaped spline (``{S-shaped}''), and unconstrained monotone spline (``{Flexible}'') models. 
		The ``{Curve},'' ``Inflection point,'' and ``50\% threshold'' columns show the root mean square error (\textsc{rmse}) and frequentist coverage probability of the 95\% credible intervals under each fitted model. 
		For both truth settings, the smallest \textsc{rmse} for each estimand is bolded.
	}
	\medskip\smallskip
		\begin{tabular}{cccccccccc}
			\hline
			Truth & Model & Knot range & \multicolumn{2}{c}{Curve} & \multicolumn{2}{c}{Inflection point} & \multicolumn{2}{c}{50\% threshold} \\ 
			\hline
			Logistic   & Logistic &                                      & $~0.16$, & \hspace{-2ex} 98\% & $~~\textbf{1.49}$, & \hspace{-3.5ex} 96\%  & $~~1.49$, & \hspace{-2.5ex} 96\%~ \\
			$(f_{\textrm{logit}})$   & S-shaped & 30 -- 90 & $~\textbf{0.14}$, & \hspace{-2ex} 69\% & $~~2.02$, & \hspace{-3.5ex} 100\% & $~~\textbf{1.09}$, & \hspace{-2.5ex} 100\%~ \\
			& S-shaped & 0 -- 120 & $~0.26$, & \hspace{-2ex} 62\% & $~~2.31$, & \hspace{-3.5ex} 100\% & $~~1.54$, & \hspace{-2.5ex} 100\% \\
			& Flexible & 30 -- 90 & $~1.07$, & \hspace{-2ex} 2\% &    &                           & $~~9.65$, & \hspace{-2.5ex} 5\% \\
			& Flexible & 0 -- 120 & $~2.56$, & \hspace{-2ex} 0.1\% &    &                           & $~~18.30$, & \hspace{-2.5ex} 14\% \\   
			\hline
			Asymmetric   & Logistic &       & $~0.21$, & \hspace{-2ex} 55\% & $~~4.46$, & \hspace{-3.5ex} 5\%  & $~~2.25$, & \hspace{-2.5ex} 77\% \\ 
			$(f_{\textrm{asym}})$   & S-shaped & 30 -- 90 & $~0.22$, & \hspace{-2ex} 55\% & $~~\textbf{3.64}$, & \hspace{-3.5ex} 63\% & $~~\textbf{1.82}$, & \hspace{-2.5ex} 79\% \\
			& S-shaped & 0 -- 120 & $~\textbf{0.20}$, & \hspace{-2ex} 69\% & $~~4.43$, & \hspace{-3.5ex} 75\% & $~~2.06$, & \hspace{-2.5ex} 95\% \\ 
			& Flexible & 30 -- 90 & $~0.91$, & \hspace{-2ex} 32\% &    &                          & $~~9.28$, & \hspace{-2.5ex} 13\% \\
			& Flexible & 0 -- 120 & $~2.32$, & \hspace{-2ex} 0.3\% &    &                          & $~~18.30$, & \hspace{-2.5ex} 23\% \\
			\hline
	\end{tabular}
	\label{sim_result}
\end{table}

Finally, we turn to more closely examining the poor performance of the unconstrained monotone spline model. 
The issue arises from the model's inability to accurately extrapolate the progression curve outside the time range spanned by observed data.
While such an extrapolation poses a challenge for any model, the parametric and S-shape models are able to better infer the curve's behavior outside the observed time window by virtue of their scientifically-motivated shape constraints.
The monotone spline without these constraints is unable, in particular, to capture the plateaus at the early and late stages of the curve. 
And the monotonicity constraint, despite its scientifically sound intention, introduces unintended upward biases by incorrectly assuming increasing trends where the curve is actually flat.
We provide in the supplement Section \ref{flex_low_perform} further discussions, along with visual explanations, of how the unconstrained monotone spline can fit the observed data and at the same time incorrectly estimate the progression curve.

\section{Real Data Application}
\label{s:result}

In this section, we apply our Bayesian hierarchical shape-constrained regression model of Section~\ref{s:method} to the longitudinal observational data from a \biocard{} study.
We investigate the temporal ordering and turning points of key \adAbbrev{} biomarkers from three categories of sources: cerebrospinal fluid (\csf{}), magnetic resonance imaging (\mri{}), and cognitive tests. 
These biomarker categories are believed to reflect different aspects of the spectrum of disease, such as chemical imbalance, damage to the brain structure, and manifested symptoms. 
For example, the deposits of amyloid beta (A$\beta$) protein in \csf{} are considered to play a central role of \adAbbrev{} progression (\citealp{jack2010}).
The decreased hippocampus volume shown on \mri{} scans reflects brain damage potentially attributable to \adAbbrev{} ~(\citealp{chandra2019magnetic}). 
Memory and visual perception measured by neurological tests are widely used for clinical diagnosis of \adAbbrev{} (\citealp{Huff1987cognitive, Jaeger2018}).
Better quantification of temporal ordering and change in these biomarkers can enhance our understanding of their relations to \adAbbrev{} progression as well as pathophysiology of the disease. 

\subsection{Data description}
\label{data-desc}

Here we describe the aspects of the \biocard{} data most essential to our analysis; 
for further details on the study and each variable, we refer interested readers to \cite{Albert2014}.

The data include 349 study participants, cognitively normal at the time of enrollment with mean age of 57.1.
Demographic variables are collected at baseline, including age, sex, years of education, and genetic markers such as Apolipoprotein E $\varepsilon$4 (\apoe{}) carrier status. 
The \csf{}, \mri{}, and cognitive biomarkers are measured approximately annually.
The study participants also received annual diagnoses, based on consensus among experts, of their cognitive statuses as either cognitively normal, mildly cognitively impaired, or having dementia. To ensure our analysis focusing on late-onset \adAbbrev{}, we excluded 46 individuals enrolled before age 50, as early-onset \adAbbrev{} differs significantly from the more common late-onset counterpart in both underlying pathophysiology and clinical manifestations \citep{koedam2010early}.
We also exclude participants with missing baseline covariates, resulting in a final sample of 253 participants with a total of 2881 visits for analysis. 

Our analysis examines the progressions of eleven biomarkers: three \csf{}, five \mri{}, and three cognitive.
The \csf{} biomarkers measure the levels of phosphorylated tau protein and of all tau proteins in aggregate, as well as the ratio of $A\beta_\textrm{1-42}$ to $A\beta_\textrm{1-40}$ proteins.
The \mri{} biomarkers include entorhinal cortex thickness, entorhinal cortex volume, and hippocampus volume, which have been averaged across the left and right hemisphere and been standardized by intracranial volume. 
The two additional \mri{} biomarkers are composite markers: the spatial pattern of abnormality for recognition of early Alzheimer’s Disease (\textsc{spare-ad}) score and the medial temporal lobe composite score. 
The cognitive markers include scores from the digit symbol substitution test, 
the logical memory test, 
and the mini mental state examination.

Not all the biomarkers are measured at every follow-up visits, with the \csf{} and \mri{} ones measured less frequently than cognitive ones since the former are more invasive and costly to collect.
The \csf{} and \mri{} are collected at $54.3\%$ of all visits, with the exception of entorhinal cortex thickness, collected only at $44.9\%$ of all visits.
On the other hand, the digit symbol substitution, logical memory, and mini mental state tests are administered at $96.4\%$, $86.5\%$, and $99.4\%$ of all visits, respectively.

The cognitive tests are preprocessed to account for learning effects, a phenomenon in which patients' performance appears to improve during the first few years of enrollment due to gradual familiarization with the test environment, even though their actual cognitive abilities can only decline over time (\citealp{Oliveira2014}). 
This artificial temporary improvement in the individual-level cognitive biomarkers, if left unadjusted, can induce bias in the subsequent estimation of the population-level curve.
Details on the preprocessing to correct for the learning effects is provided in the supplement Section~\ref{supp-learning-effect}.

Finally, we standardize biomarkers and adjust their signs so that higher values represent greater abnormality and hence more severe disease symptoms. 
The continuous baseline variables are similarly standardized.

\subsection{Analysis results}\label{model-result}

\begin{figure}[htb]
\centering
	\includegraphics[width=.95\linewidth]{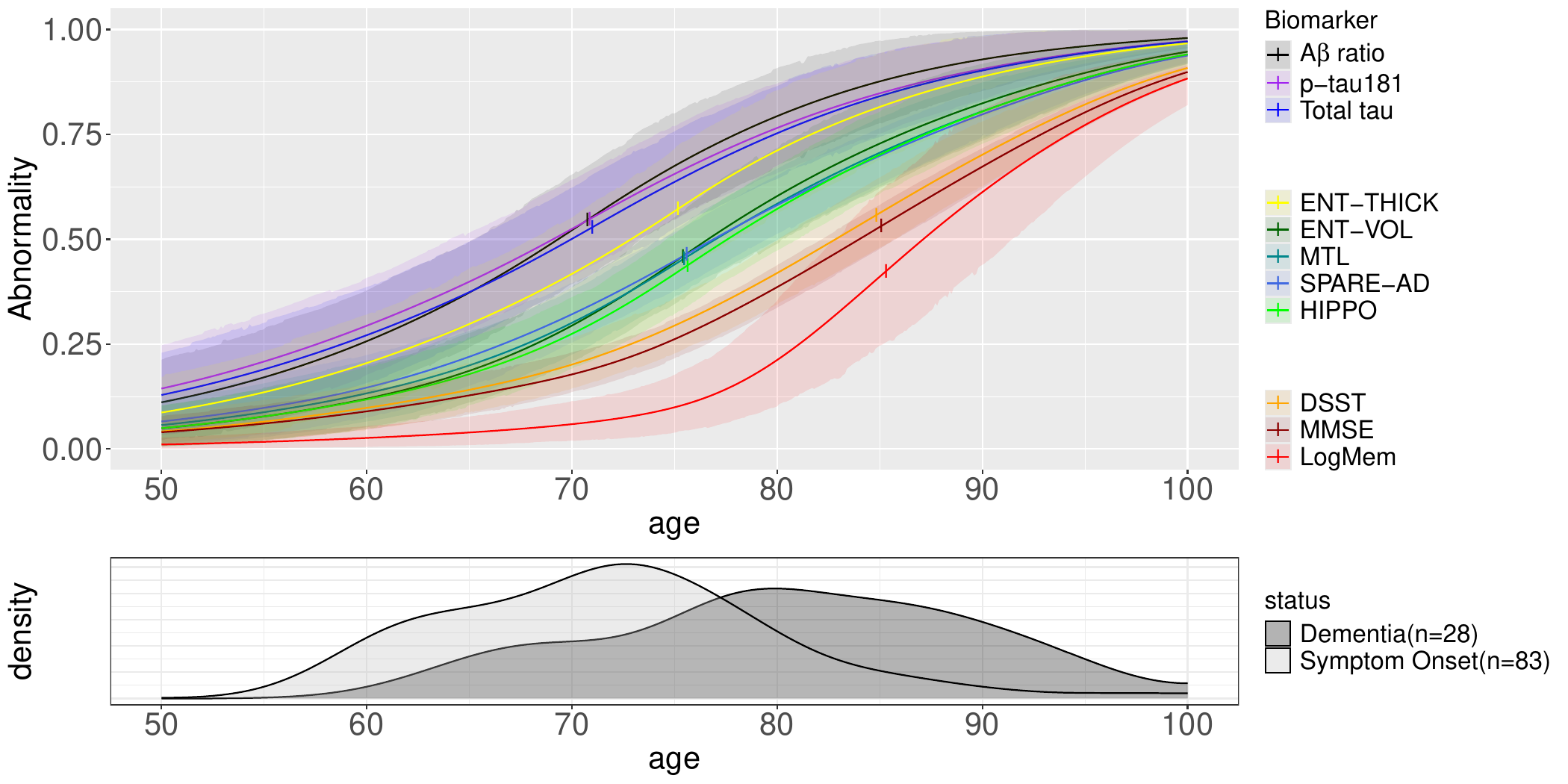}
	\caption{
		Posterior estimates along with point-wise 95\% credible intervals of standardized biomarker progression curves. 
		The vertical bars indicate the posterior means of inflection points. 
		Shown underneath are the distributions of age at first diagnosis of cognitive impairment (``Symptom Onset'') and of dementia (``Dementia'') for the 83 and 28 study participants who developed symptoms during follow-up.\\[2pt]
		\textit{Abbreviations.} The legend labels of biomarkers use the following abbreviations: 
		p-tau181~(phosphorylated tau 181);
		ENT-THICK~(Entorhinal cortex thickness); ENT-VOL~(entorhinal cortex volume); MTL~(\mri{} standardized residual composites); SPARE-AD~(spatial pattern of abnormality for recognition of early Alzheimer’s disease); HIPPO~(hippocampus volume);
		DSST~(digit symbol substitution test); 
		MMSE~(mini mental state examination); LogMem~(logical memory test).
	}
	\label{spline-band}
\end{figure}

Figure \ref{spline-band} shows the estimated population-level progression curves, along with 95\% credible intervals, of the eleven biomarkers in the standardized scale of (Equation~\ref{eq:biomarker_standardization}).
With its semi-parametric flexibility, our shape-constrained model is able to capture the asymmetry in the biomarkers' progression rates, as reflected in the gaps between the curves' inflection points and the 50\% thresholds.
The distributions of age at first diagnosis of cognitive impairment and of dementia, shown underneath the plot of progression curves, illustrate the overlap between the biomarker progressions and symptom manifestations.

The estimated progression curves suggest a clear temporal ordering among the three biomarker categories (\csf{}, \mri{}, and cognitive), 
with both 50\% thresholds and inflection points indicating the same ordering. 
The figure shows \csf{} biomarkers to reveal abnormality first, followed by \mri{}, and lastly by cognitive ones, consistent with existing literature. 
The \csf{} biomarkers start showing abnormality well before symptom onset and pass their 50\% thresholds and inflection points before the modal age of symptom onset. 
In comparison, cognitive biomarkers' remain relatively low at symptom onset, and reach maximum progression after the modal age of dementia diagnosis.

Table \ref{fixed-effect-tbl} presents the estimated effects, on the standardized biomarkers, of the baseline covariates as well as of aging from 50 to 90 years old. 
It shows that the age effect, despite being consistently the strongest, varies considerably across the biomarkers.
\Apoe{} carrier status is significantly associated with worse (i.e.\ higher) \csf{} biomarker values, and being female and having more education with better cognitive but generally worse \mri{} biomarker values.
This inverse relationship between cognitive and \mri{} values may be related to the concept of cognitive resilience, where individuals can maintain cognitive function despite suffering more \adAbbrev{}-related brain damage (\cite{Bowles2019}). 

\begin{table}[htb]
	\centering
	\begin{adjustbox}{width=\textwidth}
	\renewcommand{\arraystretch}{1.15}
	\begin{tabular}{c|c|c|c|c}
		& & & \multirowcell{2}{Education \\ (High School $\to$ College)} & \\
		& \apoe{} & Sex (Female) &  & Age $(50\to 90~\text{years})$\\ 
		\hline
		$A\beta$ ratio &  {0.77 ( 0.53, 1.01)} & \textcolor{mediumgray}{$-$0.02 ($-$0.25, 0.20)} & \textcolor{mediumgray}{$-$0.09 ($-$0.29, 0.11)} &  {1.04 ( 0.84, 1.23)}\\
		p-tau181 &  {0.44 ( 0.22, 0.68)} &  \textcolor{mediumgray}{0.15 ($-$0.09, 0.37)} & \textcolor{mediumgray}{$-$0.04 ($-$0.23, 0.16)} &  {1.56 ( 1.35, 1.77)}\\
		total tau &  {0.33 ( 0.11, 0.57)} &  \textcolor{mediumgray}{0.07 ($-$0.16, 0.30)} & \textcolor{mediumgray}{$-$0.13 ($-$0.32, 0.07)} &  {1.43 ( 1.23, 1.64)}\\  
		\hline
		ENT-THICK &  \textcolor{mediumgray}{0.04 ($-$0.21, 0.26)} &  {0.31 ( 0.09, 0.54)} &  {0.22 ( 0.02, 0.42)} &  {1.09 ( 0.73, 1.42)}\\
		ENT-VOL &  \textcolor{mediumgray}{0.04 ($-$0.20, 0.28)} &  {0.25 ( 0.02, 0.48)} &  \textcolor{mediumgray}{0.12 ($-$0.09, 0.32)} &  {1.27 ( 1.08, 1.47)}\\
		MTL &  \textcolor{mediumgray}{0.06 ($-$0.17, 0.31)} &  {0.44 ( 0.20, 0.67)} &  {0.23 ( 0.04, 0.44)} &  {1.73 ( 1.53, 1.92)}\\ 
		SPARE-AD&  \textcolor{mediumgray}{0.14 ($-$0.09, 0.37)} &  \textcolor{mediumgray}{0.20 ($-$0.05, 0.42)} &  {0.24 ( 0.03, 0.44)} &  {2.23 ( 2.04, 2.42)}\\ 
		HIPPO &  \textcolor{mediumgray}{0.10 ($-$0.14, 0.33)} & {$-$0.34 ($-$0.57, $-$0.09)} &  {0.27 ( 0.07, 0.47)} &  {1.35 ( 1.16, 1.55)}\\ 
		\hline
		DSST &  \textcolor{mediumgray}{0.03 ($-$0.18, 0.25)} & {$-$0.48 ($-$0.68, $-$0.25)} & {$-$0.28 ($-$0.47, $-$0.10)} &  {1.49 ( 1.33, 1.65)}\\ 
		MMSE &  \textcolor{mediumgray}{0.12 ($-$0.09, 0.33)} & \textcolor{mediumgray}{$-$0.07 ($-$0.28, 0.16)} & \textcolor{mediumgray}{$-$0.12 ($-$0.30, 0.07)} &  {2.36 ( 2.21, 2.51)}\\
		LogMem &  \textcolor{mediumgray}{0.11 ($-$0.10, 0.33)} & {$-$0.50 ($-$0.72, $-$0.28)} & {$-$0.30 ($-$0.49, $-$0.12)} &  {0.41 ( 0.27, 0.55)}
	\end{tabular}
	\end{adjustbox}
	\caption{
			Estimated effects---posterior mean and 95\% credible intervals (2.5\% and 97.5\% posterior quantiles)---of \apoe{} genetic marker, sex, years of education, and age on the standardized population-level biomarker progression.
			Grayed out are the estimates whose 95\% intervals include zero. 
			We show the estimated effect of going from high school (12 years) to college (16 years) graduates for the years of education, and of going from 50 to 90 years old for age.
			The biomarkers are labeled using the same abbreviations as described in the caption for Figure~\ref{spline-band}.
		}
   \label{fixed-effect-tbl}
\end{table}

\begin{figure}[htb]
\centering
	\includegraphics[width=.85\linewidth]{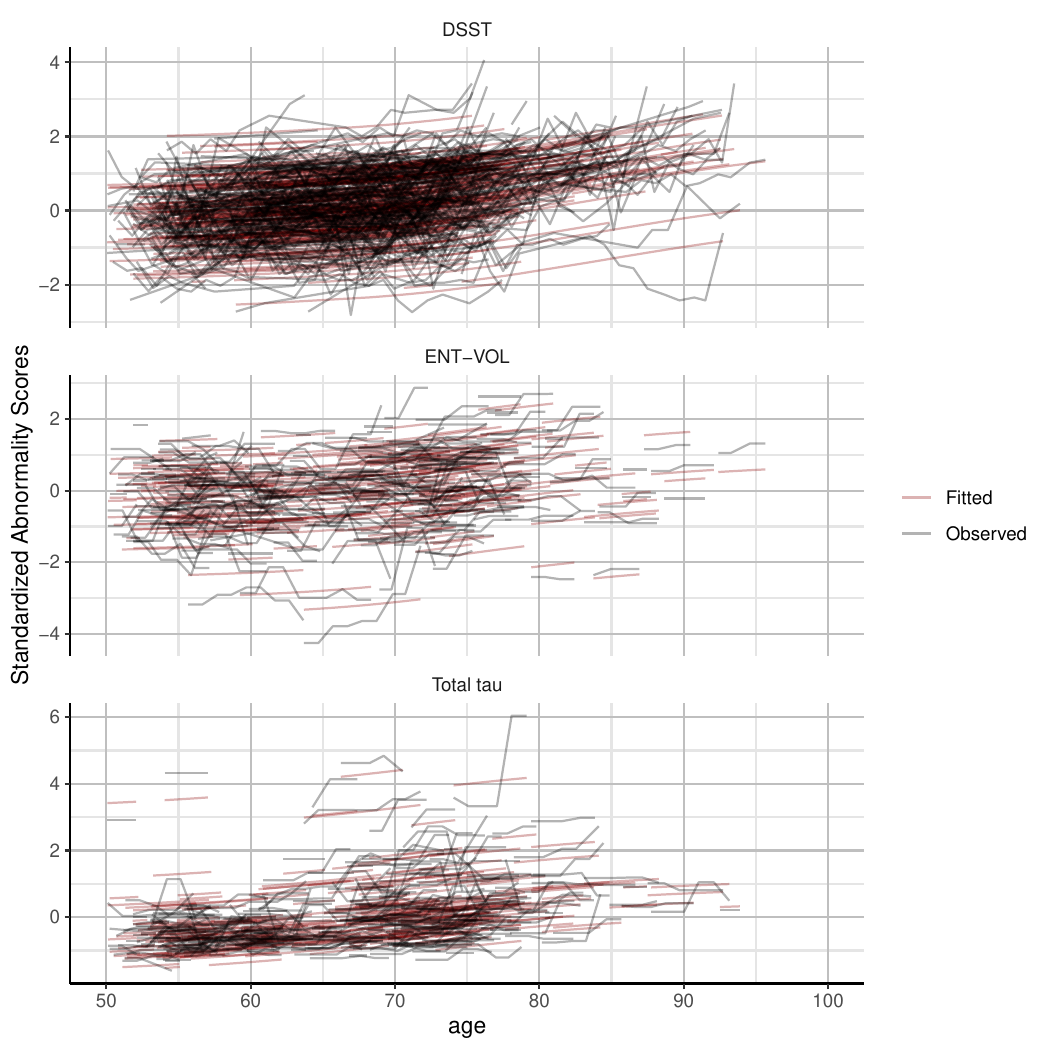}
	\caption{
		Spaghetti plots of the longitudinal progressions of selected biomarkers in the standardized scale.
		The observed values are shown in black, and the fitted values in red.
	}\label{gof-plot}
\end{figure}

Finally, we examine the model fit by comparing the observed and fitted biomarker values plotted against age in Figure~\ref{gof-plot}, where we have chosen one representative biomarker from each of the three categories for illustration.  
The fitted values are calculated as the posterior means that include the population-level curve $\hat f_k(t)$, fixed effect $x_{i}\hat\beta_k$, and subject-level intercepts $\hat\omega_{ik}$ as in Equation~\eqref{mean-model}.
The figure illustrates the high between-subject heterogeneity in baseline biomarker values, 
highlighting the importance of the subject-level random effects in our model. 
In fact, the estimated between-subject variance $\sigma_\randomeffect^2$ is $0.659$ ($95\%$ credible interval: $[0.618, 0.698]$), while the estimated residual variance $\sigma_\observed^2$ is only $0.258$ ($95\%$ credible interval: $[0.253, 0.263]$). 
The subject-level effects allow our model to fit the data reasonably despite the heterogeneity.

\FloatBarrier
\section{Discussion}
\label{s:discuss}

In this article, we have proposed a Bayesian shape-constrained regression model to quantify the progression of \adAbbrev{} biomarkers as a function of age.
The model incorporate the known structures from existing medical literature to improve model efficiency, 
while providing the semi-parametric flexibility.
The model is capable of capturing potential asymmetry in the biomarker progressions at earlier and later stages of the disease and
of providing more accurate estimates of the milestones in their progressions. 
This information can assist clinicians in diagnosing and intervening on \adAbbrev{} before irreversible damages. 

The proposed model has potential applications beyond \adAbbrev{} research since S-shaped trajectories are found in other disorders. 
Examples include the progression of motor symptoms in Parkinson’s disease and of motor, cognitive, and psychiatric symptoms in Huntington’s disease \citep{vu2012progression, koval2022forecasting, long2014tracking}.
These markers have been observed to follow S-shape trajectories, each with an initial phase of slow progression, a period of rapid progression, and then an eventual plateau. 
Our shape-constrained model could be applied, with appropriate modifications, to study the progressions of these diseases and improve our clinical understanding.

As a future research direction in studying disease biomarkers, it would be worthwhile to consider modeling the heterogeneity across individuals not only in their baseline biomarker values but also in their disease onsets and rates of disease progressions.
For the \adAbbrev{} biomarkers, in particular, Figure~\ref{gof-plot} shows a large amount of variation in the age of symptom onsets.
One way to model these forms of heterogeneity is to introduce a latent process representing summary disease progression status, building on the idea of ``distance traveled along the pathophysiological pathway’’ in \citet{Jack2013}, and model the biomarker trajectories as a function of such latent process.
We can then account for the subject-level variation in disease onsets and progression rates through subject-level intercepts and slopes for the latent disease process. 
The model for the latent process can also incorporate fixed effects; 
in fact, \apoe{} genes have been hypothesized to shift time of disease onsets \citep{Jack2013}.
The idea of introducing a latent process has been considered under a parametric assumption and frequentist framework by \citet{Jedynak2012}; 
combining it with the flexibility of our S-shaped model could provide new clinical insights, while ensuring proper uncertainty quantification through the Bayesian framework.

\section{Code and Data Availability}
\label{data-avail}
The code for the simulation study of Section \ref{s:simulation} is available in a GitHub repository at \url{https://github.com/LMY99/bayesian-s-shape-simulation-code},
and the code for the real data analysis of Section \ref{s:result} at \url{https://github.com/LMY99/biocard-analysis-code}. 
The latter repository also contains a synthetic dataset with the same format as the preprocessed \biocard{} dataset used in our analysis. 
An access to \biocard{} data can be requested at \url{https://biocard.pathology.jhu.edu/resources-for-researchers/}.

\section{Acknowledgments}
\label{s:Acknowledgments}
 This work is partially supported by the National Institutes of Health grant R01AG068002. The content is solely the responsibility of the authors and does not necessarily represent the official views of the National Institutes of Health.

\bibliographystyle{biorefs}
\bibliography{Citations}

\newpage

\renewcommand{\thesection}{S\arabic{section}}
\renewcommand{\thetable}{S\arabic{table}}
\renewcommand{\thefigure}{S\arabic{figure}}
\renewcommand{\theequation}{S\arabic{equation}}

\setcounter{section}{0}
\setcounter{figure}{0}
\setcounter{table}{0}
\setcounter{equation}{0}
\setcounter{page}{1}

\begin{center}
\fontsize{16}{24} \selectfont \bfseries
Supplement to\\
``Bayesian shape-constrained regression for quantifying Alzheimer's disease progression''
\end{center}

\markboth
{M. Li and others}
{Supplement to ``Bayesian shape-constrained regression for Alzheimer's''}

\section{Beta Kernel Density Estimation}\label{beta-kde}

Here we describe an approach to smoothing the empirical age distributions from the \adAbbrev{} data; 
as discussed in Section~\ref{ss:model}, such smoothing is required for choosing the spline knot locations that reflect the amount of available information for each spline coefficient and at the same time covers the entire span of disease progression.

Smoothing of an empirical distribution over a fixed interval can be conveniently achieved through a beta density kernel $K_\mathrm{b}(s; \alpha, \beta) = B(\alpha, \beta)^{-1} s^{(\alpha - 1)} (1 - s)^{\beta-1}$.
Specifically, given the empirical age distribution $\{t_{ij}\}_{i,j}$, we define a smoothed distribution $\hat{\rho}(t)$ on $t \in [0, 120]$ as
$$
\hat \rho(t)=\frac{1}{120 \, J} \sum_{i,j} K_\mathrm{b}\left(\frac{t}{120} \, ; \, \alpha(t_{ij}),\beta(t_{ij})\right),
$$
where $J = \sum_i J_i$ is the total number of observed time points and the shape parameters $\alpha(s)$ and $\beta(s)$ are chosen in a location-specific manner as
\begin{equation}\label{beta-kde-shape}
	\alpha(s)=\nu s+1, \ \beta(t)=\nu(1-s)+1
\end{equation}
for $\nu > 0$.
The above choice of $\alpha(s)$ and $\beta(s)$ ensure that the kernel $K_\mathrm{b}(t ; \alpha(t_{ij}), \beta(t_{ij}))$ has its mode as $t = t_{ij}$.
The parameter $\nu$ controls the kernel's variance, with smaller values corresponding to larger variance and flatter estimates of $\hat \rho(t)$. 
We use $\nu=10$ in both our simulation study of Section~\ref{s:simulation} and real data analysis of Section~\ref{s:result}.

\section{Details on Markov chain Monte Carlo of Section~\ref{MCMC}}\label{mcmc-detail}

The posterior for $\beta,\gamma,\omega,\sigma_\observed^2,\sigma_\randomeffect^2,\sigma_\smooth^2,\sigma_\vanishderiv^2$ is given by the Bayes' rule as 
\begin{multline*}
p(\beta,\gamma,\omega,\sigma_\observed^2,\sigma_\randomeffect^2,\sigma_\smooth^2,\sigma_\vanishderiv^2
	\, | \,  x,y,t) \\
	\propto p(\sigma_\smooth^2,\sigma_\vanishderiv^2)p(\sigma_\observed^2)p(\sigma_\randomeffect^2)p(\beta)p(\omega \, | \, \sigma_\randomeffect^2)p(\gamma \, | \, \sigma_\smooth^2,\sigma_\vanishderiv^2)p(y \, | \, \beta,\gamma,x,t,\omega,\sigma_\observed^2),
\end{multline*}
with the likelihood and priors given according to the specifications of Sections~\ref{ss:hetero} and \ref{ss:model} as follows.
The priors for $\beta,\omega,\sigma_\observed^2,\sigma_\randomeffect^2,\sigma_\smooth^2,\sigma_\vanishderiv^2$ are given as
	$$
	\begin{aligned}
	p(\sigma_\smooth^2,\sigma_\vanishderiv^2)
		&\propto\exp\left[-\frac{\sigma_\smooth^2}{2\cdot (1/20)^2}-\frac{\sigma_\vanishderiv^2}{2\cdot (1/20)^2}\right]\\
	p(\sigma_\observed^2)&\propto (\sigma_\observed^2)^{-4}e^{-0.5/\sigma_\observed^2}\\
	p(\sigma_\randomeffect^2)&\propto (\sigma_\randomeffect^2)^{-4}e^{-0.5/\sigma_\randomeffect^2}\\
	p(\beta)&\propto \prod\limits_{k=1}^K \exp\left(-\frac{1}{2\cdot 100^2}\beta_k'\beta_k\right)\\
	p(\omega \, | \, \sigma_\randomeffect^2)&\propto \prod\limits_{k=1}^K\prod\limits_{i=1}^N\left(\sigma_\randomeffect ^2\right)^{-1/2}\exp\left(-\frac{1}{2\sigma_\randomeffect^2}\omega_{ik}^2\right).
	\end{aligned}
	$$
The prior for $\gamma$ is given by
	\begin{align*}
	p(\gamma \, | \, \sigma_\smooth^2,\sigma_\vanishderiv^2)
		&\propto \frac{1}{Z\!\left(\sigma_\smooth^2,\sigma_\vanishderiv^2, {\Gamma_{\mathrm{grp}}} \right)} \prod\limits_{k=1}^K\exp\left[-\frac{1}{2}\gamma_k'V_{\gamma }^{-1}(\sigma_\smooth^2,\sigma_\vanishderiv^2)\gamma_k\right]\cdot \indFun\left[\left(\gamma_1,\cdots,\gamma_K\right)\in\Gamma_{\mathrm{grp}}\right],
	\end{align*}
	where $V_\gamma(\sigma_\smooth^2,\sigma_\vanishderiv^2)$ is a $\numSplines\times \numSplines$ symmetric matrix such that
	$$
	\gamma'V_\gamma^{-1}(\sigma_\smooth^2,\sigma_\vanishderiv^2)\gamma=\sum\limits_{\splineIndex=1}^{\numSplines}\frac{(\gamma_{\splineIndex}-\gamma_{\splineIndex-1})^2}{\sigma_\smooth^2}-\sum\limits_{\splineIndex=1}^{\numSplines}
	\frac{\gamma_{\splineIndex}^2}{\alpha_{\numSplines}(\splineIndex)\sigma_\vanishderiv^2},
	$$
	and $\Gamma_{\mathrm{grp}}$ denotes the set in which $\gamma_1,\gamma_2,\cdots,\gamma_K$ satisfy the linear constraints (\ref{eq:prior_constraint}) and the coefficients within each of the three biomarker groups share the same inflection index.
	And $Z\!\left(\sigma_\smooth^2,\sigma_\vanishderiv^2 \right)$ is part of the normalizing constant, representing an integral of Gaussian density over $\Gamma_{\mathrm{grp}}$:
	\begin{equation}
	\label{eq:trunc_gaussian_norm_const}
	Z\!\left(\sigma_\smooth^2,\sigma_\vanishderiv^2, {\Gamma_{\mathrm{grp}}} \right) 
		= P\left\{(
			\tilde{\gamma}_1,\cdots,\tilde{\gamma}_K)\in\Gamma_{\mathrm{grp}} \mid \tilde{\gamma}_k \sim  \normalDist(0,V_{\gamma}(\sigma_\smooth^2,\sigma_\vanishderiv^2)) \text{ for each } k \,
		\right\}.
	\end{equation}
	Finally, the likelihood is given by
	\begin{multline*}
	p(y \, | \, \beta,\gamma,x,t,\omega,\sigma_\observed^2)\\
		\propto\prod\limits_{i=1}^N\prod\limits_{j=1}^{J_i}\prod\limits_{k=1}^K(\sigma_\observed^2)^{-1/2}\exp\left\{-\frac{1}{2\sigma_\observed^2}\left[y_{ijk}-x_{i}'\beta_k-\sum\limits_{\splineIndex=1}^{\numSplines}\gamma_{k\splineIndex}I_{\splineIndex}(t_{ij})-\omega_{ik}\right]^2\right\}.
	\end{multline*}
	
	We first focus our discussion on the parameters whose conditional distributions do not belong to standard parametric families and hence require advanced sampling techniques.
	The full conditional density of $(\beta,\gamma)$ is given as
	$$
	\begin{aligned}
		&p(\beta,\gamma \, | \, \omega,\sigma_\observed^2,\sigma_\randomeffect^2,\sigma_\smooth^2,\sigma_\vanishderiv^2,x,y,t)\\
		&\hspace*{2em}\propto\prod\limits_{k=1}^K\prod\limits_{i=1}^N\prod\limits_{j=1}^{J_i}\exp\left[-\frac{1}{2\sigma_\observed^2}\left(y_{ijk}-x_{i}'\beta_k-\sum\limits_{\splineIndex=1}^{20}\gamma_{k\splineIndex}I_{\splineIndex}(t_{ij})-\omega_{ik}\right)^2\right]\\
		&\hspace*{8em}\cdot\exp\left[-\frac{1}{2}\begin{pmatrix}\beta_k\\\gamma_k\end{pmatrix}'
		\begin{pmatrix}100^2 \mathrm{I}&0\\0&V_{\gamma}(\sigma_\smooth^2,\sigma_\vanishderiv^2)\end{pmatrix}^{-1}
		\begin{pmatrix}\beta_k\\\gamma_k\end{pmatrix}\right]
		\cdot \indFun[(\gamma_1,\cdots,\gamma_K)\in\Gamma_{\mathrm{grp}}],
	\end{aligned}
	$$
	which is a truncated version of multivariate Gaussian distribution with variance
	$$
	V^*_{k}=\left[\begin{pmatrix}100^2 \mathrm{I}&0\\0&V_{\gamma }(\sigma_\smooth^2,\sigma_\vanishderiv^2)\end{pmatrix}^{-1}+\sigma_\observed^{-2}\sum\limits_{i=1}^N\sum\limits_{j=1}^{J_i}\begin{pmatrix}x_{i}\\I(t_{ij})\end{pmatrix}\begin{pmatrix}x_{i}\\I(t_{ij})\end{pmatrix}'\right]^{-1}
	$$
	and mean
	$$
	\mu^*_{k}=V^*_{k}\left[\sigma_\observed^{-2}\sum\limits_{i=1}^N\sum\limits_{j=1}^{J_i}\begin{pmatrix}x_{i}\\I(t_{ij})\end{pmatrix}(y_{ijk}-\omega_{ik})\right],
	$$
	where $I(t)=(I_{1}(t),\cdots,I_{\numSplines}(t))$ denotes a vector of the basis function values at $t$.
	To sample from this truncated Gaussian distribution, we first sample the shared inflection index $\splineIndex_g^*$ for each of the three  biomarker groups.
	The indices are conditionally independent with the full conditional distribution
	$$
	P(\splineIndex_g^*=\splineIndex \, | \, \dots )
		\propto \prod_{k \in \mathcal{C}_g} P\left\{
			(\tilde{\beta}_k, \tilde{\gamma}_k)\in\mathbb{R}^q\times\Gamma_\splineIndex \, | \, (\tilde{\beta}_k, \tilde{\gamma}_k)\sim \normalDist(\mu_k^*,V_{k}^*)
		\right\},
	$$
	where $\mathcal{C}_g$ denotes the set of biomarker indices belonging to the $g$-th category and $\Gamma_\splineIndex$ is the constrained region as defined in Equation~\eqref{eq:prior_constraint}. 
	The parameters $(\beta_k,\gamma_k)$ for $k = 1, \ldots, K$ are independent conditionally on these inflection indices;
	we sample each of them from $\normalDist(\mu_k,V_k^*)$ truncated on $\mathbb{R}^q\times\Gamma_{\splineIndex_g^*}$. 
	
	The full conditional of $(\sigma_\smooth^2,\sigma_\vanishderiv^2)$ is given as
	\begin{align*}
	&p(\sigma_\smooth^2,\sigma_\vanishderiv^2 \, | \, \beta,\gamma,\omega,\sigma_\observed^2,\sigma_\randomeffect^2,x,y,t) \\
	&\propto
			\exp\left[-\frac{\sigma_\smooth^2}{2 (1/20)^2}-\frac{\sigma_\vanishderiv^2}{2 (1/20)^2}\right]
			\frac{1}{Z\!\left(\sigma_\smooth^2,\sigma_\vanishderiv^2, {\Gamma_{\mathrm{grp}}} \right)} \prod\limits_{k=1}^K\left\{\det\left(V_\gamma^{-1}\right)\exp\left[-\frac{1}{2}\gamma_k'V_{\gamma}^{-1}(\sigma_\smooth^2,\sigma_\vanishderiv^2)\gamma_k\right]\right\},
	\end{align*}
	where $Z\!\left(\sigma_\smooth^2,\sigma_\vanishderiv^2, \Gamma_{\mathrm{grp}} \right)$ is the integral of Gaussian density over $\Gamma_{\mathrm{grp}}$ as in Equation~\eqref{eq:trunc_gaussian_norm_const}.
	We update the two parameters jointly through the Metropolis algorithm. 
	For the required evaluation of the Gaussian density integral, we note that 
	$Z\!\left(\sigma_\smooth^2,\sigma_\vanishderiv^2, \Gamma_{\mathrm{grp}} \right) 
		= \prod\limits_{g=1}^3\sum\limits_{\splineIndex^*=1}^{\numSplines}\prod\limits_{k \in \mathcal{C}_g} Z(\sigma_\smooth^2,\sigma_\vanishderiv^2, \Gamma_{m^*})$
	where
	$$ Z(\sigma_\smooth^2,\sigma_\vanishderiv^2, \Gamma_{\splineIndex^*}) 
		= P\left\{ 
			\tilde{\gamma}_k\in\Gamma_{\splineIndex^*} \, | \, \tilde{\gamma}_k \sim \normalDist(0,V_{\gamma}(\sigma_\smooth^2,\sigma_\vanishderiv^2)) 
		\right\}, $$
	which we compute with the method of \citet{Genz1992}.
	We carry out the update in log-scale, with Gaussian proposals. 
	The proposal variance is tuned through the adaptive Metropolis algorithm of \citetsupplement{Roberts2006}.
    
    For the rest of the parameters, their full conditional distributions belong to standard parametric families and can be straightforwardly sampled from. 
    In summary, the Gibbs sampler cycles through following steps:

	\begin{itemize}
		\item Sample $(\beta_k,\gamma_k)$ from its conditional using the strategy described above.
	
		\item Sample $\sigma_\observed^2$ from inverse-gamma distribution with shape $3 + (J_\mathrm{tot} / 2)$ and scale \\ $0.5 + \frac{1}{2}\sum_{i,j,k}
			\left( y_{ijk}-x_{i}'\beta_k-\sum_{\splineIndex=1}^{20}\gamma_{k\splineIndex}I_{\splineIndex}(t_{ij})-\omega_{ik} \right)^2$ where $J_\mathrm{tot}$ is the total number of observed biomarker entries.
		\item Sample $\sigma_\randomeffect^2$ from inverse-gamma with shape $3+KN$
		and scale $0.5+\frac{1}{2}\sum\limits_{i=1}^N\sum\limits_{k=1}^K\omega_{ik}^2$.
		\item Sample $\omega_{ik}$ from a Gaussian distribution with variance $\left( \frac{1}{\sigma_\randomeffect^2}+\frac{J_i}{\sigma_\observed^2} \right)^{-1}$
		and mean \\ $\frac{1/\sigma_\observed^2}{J_i/\sigma_\observed^2+1/\sigma_\randomeffect^2}\sum\limits_{j=1}^{J_i}(y_{ijk}-x_{i}'\beta_k-\sum\limits_{\splineIndex=1}^{20}\gamma_{k\splineIndex}I_{\splineIndex}(t_{ij}))$.
		\item Sample $(\sigma_\smooth^2,\sigma_\vanishderiv^2)$ via Metropolis from its conditional using the strategy described above.
	\end{itemize}
	
	\section{Underperformance of the Flexible Spline Model in Section~\ref{sim:results}}\label{flex_low_perform}

\begin{figure}[h]
		\includegraphics[width=\linewidth, page=1]{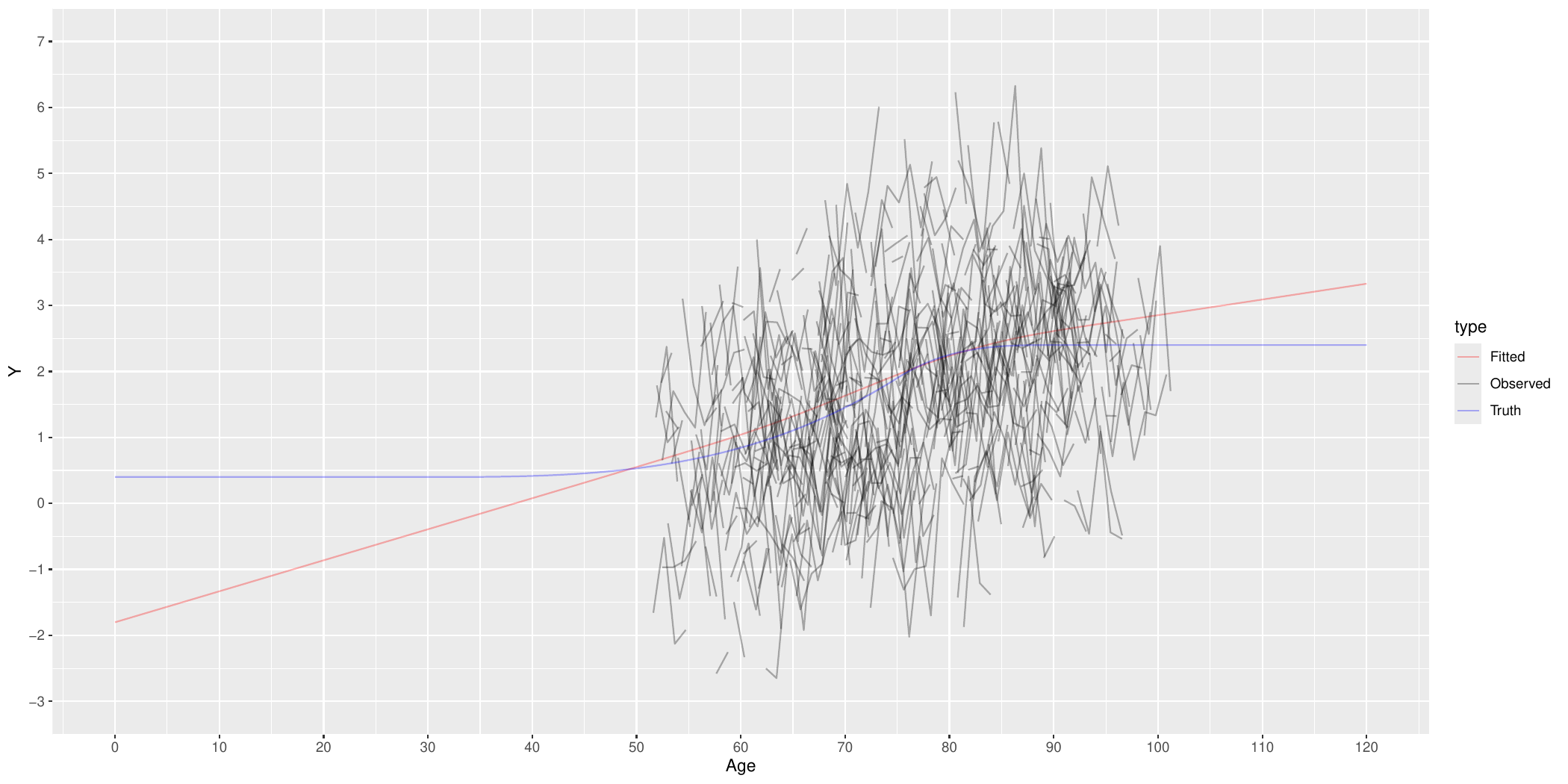}
		\caption{
			Spaghetti plot of the observed biomarkers along with the true population-level progression curve (blue) and the estimated curve from the unconstrained monotone spline model (red). 
			The dataset used coincides with one of our 100 synthetic datasets under asymmetric truth $f_{\textrm{asym}}$ from Section~\ref{s:simulation}.
		}
		\label{flex_diagnosis_Y}
	\end{figure}

	As we observed in Section~\ref{sim:results}, the unconstrained monotone spline model performs poorly in estimating the population-level progression curve.
	This is due to the fact that the monotonicity constraint necessarily forces the prior means of spline coefficients to be positive and thereby biases the posterior estimates upward, especially in the absence of the vanishing derivative constraint.
	The estimates of subject-level curves can still fit the data reasonably with the overall and subject-level intercepts compensating for the bias in the population-level curve, as seen in Figure~\ref{flex_diagnosis_Y}.
	
	\section{Adjustment for Learning Effects in Cognitive Tests}\label{supp-learning-effect}
	
	Here we describe how we preprocess the biomarker data from cognitive tests to account for the phenomenon of learning effects discussed in Section~\ref{data-desc}.
	Based on existing literature and exploratory analysis of the \biocard{} data, we assume the learning to take place over patients' first three annual visits and then reaches a steady state afterward.
	We therefore assume the learning effects' influence on the $k$-th biomarker, if it belongs to the cognitive category, to take a piecewise linear function of the form
	\begin{equation*} 
		\ell_k(t ; \alpha_k) = \alpha_k \cdot \min(t,3).
	\end{equation*}
	We estimate $\alpha_k$ by taking the cognitive test scores from the visits within the first three years of enrollment and then regressing the changes in the score on the elapsed time;
	i.e. we regress $\, y_{ik}(t_{i(j + 1)}) - y_{ik}(t_{i1})$ on $\, t_{i(j + 1)} - t_{i1}$ using the data from all the visits satisfying $t_{i(j+1)}-t_{i1}\le 3$.
	Subsequently, we subtract the estimated learning effects $\ell_k(t; \hat{\alpha}_k)$ from the cognitive test scores and use these preprocessed values as outcomes in our model. 

\bibliographystylesupplement{biorefs}
\bibliographysupplement{supp}

\end{document}